\documentclass [12pt,a4paper]{article}

\usepackage {amssymb}
\usepackage {amsmath}
\usepackage{amsthm}
\def\dse#1{\vskip 0.6cm\noindent
        {\large\bf #1}
        \vskip 0.4cm}

\begin{document}
\begin{center}
\textbf{\Large{On the $k$-error linear complexity of subsequences of $d$-ary Sidel'nikov sequences over prime field  $\mathbb{F}_{d}$ } }\footnote {
}
\end{center}

\begin{center}
\small Minghui Yang$^a$, Jiejing Wen$^b$
\end{center}

\begin{center}
\textit{\footnotesize a The State Key Laboratory of Information Security, Institute of Information Engineering, Chinese Academy of Sciences, Beijing 100195,  China\\
b Chern Institute of Mathematics, Nankai University, Tianjin 300071, China}
\end{center}


\noindent\textbf{Abstract:} We study the $k$-error linear complexity of subsequences
of the $d$-ary Sidel'nikov sequences over the prime field $\mathbb{F}_{d}$. A general lower bound for the $k$-error linear complexity is given. For several special
 periods,  we show that these sequences have large $k$-error
linear complexity.

\noindent\textbf{keywords:} $k$-error linear complexity, subsequences, Sidel'nikov sequences

\dse{1~~Introduction}The linear complexity $LC(\{s_n\})$ [4, Lemma 8.2.1] of an $N$-periodic sequence $\{s_n\}=s_0, s_1,\ldots$ over the field $\mathbb{F}$
is the smallest nonnegative integer $L$ such that there exist coefficients
$c_0, c_1, \ldots,c_{L-1} \in \mathbb{F}$ such that

$$s_{n+L}+c_{L-1}s_{n+L-1}+\cdots+c_0s_{n}=0  \ \textrm {for all} \ n \geq 0.$$
and can be computed by
\begin{equation}
LC(\{s_n\})=N-\deg(\gcd(x^N-1, S(x))),
\end{equation}
 where $S(x)=s_0+s_1x+\cdots+s_{N-1}x^{N-1}. $

Linear complexity is of fundamental importance as cryptographic characteristic of sequences \cite{R} . Motivated
by security issues of stream ciphers, Stamp and Martin proposed the concept of the $k$-error linear complexity \cite{ST}.
 The $k$-error linear complexity $LC_k(\{s_n\})$ of a sequence $\{s_n\}$ over the field $\mathbb{F}$ is defined  as the smallest
linear complexity that can be obtained by changing at most $k$ terms of the sequence per period.
The concept of the $k$-error linear complexity is built on the
earlier concepts of weight complexity introduced in \cite{D} and sphere complexity introduced in \cite{DX}.

Let $q$ be a power of an odd prime $p$, $\gamma$ a primitive element of
$\mathbb{F}_{q}$, and let $d$ be a positive prime divisor of $q-1$.
 Then the cyclotomic classes of order $d$ give a partition of
$\mathbb{F}_{q}^{\ast}=\mathbb{F}_{q} \backslash \{0\}$ defined by
$$ D_0=\{\gamma^{dn}: 0 \leq n \leq (q-1)/d-1\} \ \textrm {and} \  D_j=\gamma^j D_0, 1 \leq j \leq d-1. $$

Let $l$ be a divisor of $q-1$ and $\alpha=\gamma^{(q-1)/l}$. In this paper, we investigate the
$l$-periodic sequence $\{s_n\}$ with terms in the finite field $\mathbb{F}_{d}$ defined by
\begin{equation} s_n=
\begin{cases} j & \textrm{if $\alpha^n+1 \in D_j$} \\0 & \textrm{if $\alpha^n+1=0.$}
\end{cases}
\end{equation}

For $l=q-1$, these sequences are Sidel'nikov sequences which were independently introduced by Sidel'nikov \cite{SI}
and by Lempel, Cohn and Eastman for the case $d=2$ \cite{LE}. For $l < q-1$, these sequences are obviously the
subsequences of Sidel'nikov sequences.  It is known that every  Sidel'nikov sequence has good autocorrelation properties, see \cite{K, LE}.
The linear complexity and $k$-error linear complexity of the $d$-ary Sidel'nikov sequence over $\mathbb{F}_{p}$ have been investigated
in \cite{AM,GA,H}. Using some facts in character theory, the linear complexity of the $d$-ary Sidel'nikov sequence over $\mathbb{F}_{d}$ was analyzed in \cite{BM} and the $k$-error linear complexity of subsequences
of binary Sidel'nikov sequence over $\mathbb{F}_{2}$ and $\mathbb{F}_{p}$ were considered in \cite{BW} .

In the next section, combining the methods of \cite{BM} and \cite{BW}, we prove several results on the $k$-error linear complexity of subsequences
of the $d$-ary Sidel'nikov sequences over the prime field $\mathbb{F}_{d}$. We give a general lower bound. Furthermore, for several special periods, we give the exact values of the $k$-error linear complexity. The results show that for several special cases,
these sequences are good from the viewpoint of the $k$-error linear complexity.

This paper is organized as follows. In section 2, we discuss the $k$-error linear complexity of the $d$-ary Sidel'nikov sequence over $\mathbb{F}_{d}$. We conclude this paper in section 3.

\dse{2~~The $k$-error linear complexity over $\mathbb{F}_{d}$}
 We present a general lower bound first.

 To do this, the following Lemma about Weil's bound is needed.

\par\textbf{Lemma 1} ([12, Theorem 5.41]) Let $\psi$ be a multiplicative character of $\mathbb{F}_q$ of order $m>1$ and let $f\in\mathbb{F}_q[x]$
 be a monic polynomial of positive degree that is not an $m$th power of a polynomial. Let $e$ be the number of distinct roots of $f$ in its splitting field over $\mathbb{F}_q$. Then for every $a\in\mathbb{F}_q$ we have
 $$|\sum_{c\in\mathbb{F}_q}\psi(af(c))|\leq(e-1)q^{1/2}.$$
\par\textbf{Theorem 1} The $k$-error linear complexity of the sequence $\{s_n\}$ defined in (2) over $\mathbb{F}_{d}$ satisfies
that if $l$ is odd, then $LC_k(\{s_n\})>l/(q^{1/2}+2k)-1$, otherwise $LC_k(\{s_n\})>l/(q^{1/2}+2k+2)-1$.

\begin{proof}
Let $\{t_n\}$ be a sequence with period $l$ over  $\mathbb{F}_{d}$ which is obtained by changing at most
$k$ terms of the notations defined as before, the sequence $\{s_n\}$ per period. Let $LC(t_n)=L$, $c_L=1$, then we have
\begin{equation}
t_{n+L}+c_{L-1}t_{n+L-1}+\cdots+c_0t_{n}=0 \ (n \geq 0).
\end{equation}
With the notations as before, let $\chi$ denote a nontrivial multiplicative character with $\chi(\gamma^j)=\xi_d^j$ $(0\leq j\leq q-2)$, where $\xi_d=e^{2\pi\sqrt{-1}/d}$, then we have $\xi_d^{s_n}=\chi(\alpha^n+1)$, if $l$ is odd or $l$ is even and $n\neq l/2$. In what follows, we only consider the case that
$l$ is odd. When $l$ is even, the results can be similarly proven. If $l$ is odd, we have $\xi_d^{t_n}=\chi(\alpha^n+1)$ for at least $l-k$ terms of per period of $\{t_n\}$. Then from (3), for at least $l-k(L+1)$ terms of each period of $\{t_n\}$ we have
$$\chi(\prod_{m=0}^L(\alpha^{n+m}+1)^{c_m})=\prod_{m=0}^L\xi_d^{t_{n+m}c_m}=\xi_d^{\sum_{m=0}^Lt_{n+m}c_m}=1.$$

So

\begin{equation*}\begin{split}
l-2k(L+1)& \leq |\sum_{n=0}^{l-1}\chi(\prod_{m=0}^L(\alpha^{n+m}+1)^{c_m})|\\& =\displaystyle\frac{l}{q-1} |\sum_{n=0}^{q-2}\chi(\prod_{m=0}^L(\gamma^{\frac{q-1}{l}n}\alpha^{m}+1)^{c_m})| \\
                                  & \leq\displaystyle\frac{l}{q-1} [(\frac{q-1}{l}(L+1)-1)\sqrt{q}+1]\\& <(L+1)\sqrt{q}
,
\end{split}\end{equation*}
where the penultimate step is obtained from Lemma 1, then the results are proven.
\end{proof}
Now we give lower bounds for some special periods which improve Theorem 1.

\par\textbf{Proposition 1} Let $r  (r \neq d)$ be an odd prime divisor of $l$. Let $\{t_n\}$ be a sequence obtained by altering at most
 $k$ elements of $\{s_n\}$ and $T(x)=t_0+t_1x+\cdots+t_{l-1}x^{l-1}$. From the fact that $d$ is a primitive root modulo $r$ and $r\geq\sqrt{q}+2k+1$,
then for each $r$-th root of unity $\beta\neq1$ we have $T(\beta)\neq0$.

\begin{proof}
  We prove it by contradiction. Assume that $T(\beta)$=0. As $\beta^r=1,$ we have $T(\beta)=\sum_{n=0}^{l-1}t_n\beta^n=\sum_{b=0}^{r-1}\sum_{j=0}^{l/r-1}t_{b+jr}\beta^b.$ From $d$ is a primitive root modulo $r$, we know $\Psi(x)=\sum_{b=0}^{r-1}x^b$ is the minimal polynomial of $\beta$ over $\mathbb{F}_d$. Then $\sum_{j=0}^{l/r-1}t_{jr}=\sum_{j=0}^{l/r-1}t_{1+jr}=\ldots=\sum_{j=0}^{l/r-1}t_{r-1+jr}.$

For at least $l-k-1$  many $n$ of one period of the sequence, we have
\begin{equation}
\xi_d^{t_n}=\xi_d^{s_n}=\chi(\alpha^n+1).
\end{equation}

As
$$\prod_{j=0}^{l/r-1}(\alpha^{jr}x+1)=1-(-1)^lx^{l/r},$$
combining with (4), for at least $r-k-1$ or $r-k$ many $b$ in the set $\{0, 1, \ldots, r-1\}$ if $l$ is even or odd, respectively, we have
$$\xi_d^{\sum_{j=0}^{l/r-1}t_{b+jr}}=\prod_{j=0}^{l/r-1}\chi(\alpha^{b+jr}+1)=\chi(1-(-1)^l\alpha^{bl/r})=e,$$ where $e$ is a constant.
Then
\begin{equation*} |\sum_{b=0}^{r-1}\chi(1-(-1)^l\alpha^{bl/r})|\geq
\begin{cases} r-2k & \textrm{if $l$ is odd} \\r-2k-1 & \textrm{if $l$ is even, }
\end{cases}
\end{equation*}
according to the fact that when $l$ is even and $r$ is odd, $\chi(0)$ appears in the sum only once.

So
\begin{equation*}\begin{split}
r-2k-1& \leq |\sum_{b=0}^{r-1}\chi(1-(-1)^l\alpha^{bl/r})| \\
                                  &= \displaystyle\frac{r}{q-1}|\sum_{b=0}^{q-2}\chi(1-(-1)^l\gamma^{b(q-1)/r})|<\sqrt{q}
,
\end{split}\end{equation*}
where the  penultimate step is followed by Weil's bound. This contradicts our assumption on $r$.
\end{proof}
\par\textbf{Corollary 1} Let $l=d^mrv$, where $r$ is a prime and $r\geq\sqrt{q}+2k+1$, $r, v$ are coprime with $d$ and $d$ is a primitive root modulo $r$. Then we have $LC_k(\{s_n\})\geq(r-1)d^m$.

\begin{proof}
For each $r$th root of unity $\beta\neq1$, we have $T(\beta)\neq0$ according to Proposition 1. This implies that
the polynomial $(\displaystyle\frac{x^r-1}{x-1})^{d^m}$ is coprime with $T(x)=\sum_{i=0}^{l-1}t_nx^n$. Then from (1), we have $LC_k(\{s_n\})\geq(r-1){d^m}$.
\end{proof}
Now we give exact values of the 1-error linear complexity of the sequence defined in (2) when $d=3$ for some special
cases.

If $l=d^sr$ and $gcd(d,r)=1$, then $x^l-1=(x^r-1)^{d^s}$. The Hasse derivative $S(x)^{(h)}$ is employed to determine the multiplicity
of the roots of unity for $S(x)$, which is defined to be
$$S(x)^{(h)}=\sum_{n=h}^{l-1}\binom n hs_nx^{n-h}.$$
The multiplicity of $\theta$ as a root of $S(x)$ is $u$ if it satisfies $S(\theta)=S(\theta)^{(1)}=\ldots=S(\theta)^{(u-1)}$ and
$S(\theta)^{(u)}\neq0$ ([12, Lemma 6.51]). The  binomial coefficients appearing in $S(x)^{(h)}$ can be evaluated by Lucas' congruence \cite{GR}
$$\binom nh\equiv\binom{n_0}{h_0}\cdots\binom{n_e}{h_e}\mod d $$
where $n_0, \ldots, n_e$ and $h_0, \ldots, h_e$ are the digits in the $d$-ary representation of $n$ and $h$ respectively. It
is easy to see that
\begin{equation}
\binom n h\equiv\binom i h\mod d.
\end{equation}
for $h<d^e$ and $n\equiv i\mod d^e$.

With the cyclotomic classes of order $v$ denoted by $D_j$, the cyclotomic numbers $(i,j)_v$ (see \cite{CD}) are defined by
$(i,j)_v=|(D_i+1) \cap D_j|,  0\leq i, j\leq v-1.$
We can express the $h$th Hasse derivative corresponding to the sequence defined in (2) using (5),

\begin{align}
S(1)^{(h)} &=\sum_{n=h}^{l-1}\binom n hs_n=\sum_{i=h}^{d^e-1}\binom i h\sum_{n=h\atop n\equiv i\mod d^e }^{l-1}s_n \notag \\
                                  &= \sum_{i=h}^{d^e-1}\binom i h\sum_{h=0\atop n=hd^e+i }^{l/d^e-1}\sum_{m=1}^{d-1}\sum_{s_n=m}m \notag\\
                                  &=\sum_{i=h}^{d^e-1}\binom i h\sum_{j=0}^{\frac{q-1}{l}d^{e-1}-1}\sum_{m=1}^{d-1}(\frac{q-1}{l}i, jd+m)_{\frac{q-1}{l}\cdot d^e}m
.
\end{align}

Let $q=cf+1$, the relation between the cyclotomic numbers of order $c$ is given in \cite{CD}
\begin{align}
 (i, j)_c=(c-i, j-i)_c= \left\{ \begin{array}{ll}
(j, i)_c & f \ even\\
(j+c/2, i+c/2)_c& f \ odd.
\end{array} \right.
\end{align}

We use the expressions of $S(1)^{(h)}$ to get the multiplicity of 1 as a root of $S(x)$.   If the corresponding cyclotomic numbers are known, then from the multiplicity of 1 as a root of $S(x)$ we can get the exact value of $LC_1(\{s_n\})$ for some special cases. We take $l=\frac{q-1}{2}$, $d=3$ as an example.

To get the following theorem, we need to use cyclotomic numbers of order 6 that rely on the unique decomposition  $q=6f+1=A^2+3B^2$ of $q$ with $A\equiv 1\bmod3$ and moreover $\gcd(A,q)=1$ when $q=p^m$ and $p\equiv1\bmod 6$. The sign of $B$ relies on the choice of the primitive element $\gamma$.

\par\textbf{Theorem 2} Let $\{s_n\}$ be a sequence defined in (1) over the finite field $\mathbb{F}_3$ with period $l=(q-1)/2 $, where $l=3^ar \ (a\geq 1)$, $r$ is a prime, $r\neq3$,
3 is a primitive root modulo $r$ and $r\geq\sqrt{q}+3$. If $B\equiv0 \mod 3$, then
$LC_1(\{s_n\})=LC(\{s_n\}).$
Furthermore, if
$B\equiv0 \mod 3$ and $A\not\equiv1\mod 9$ then
$LC_1(\{s_n\})=l-1.$

\begin{proof}
From (6), we have \begin{equation*}\begin{split}
S(1)^{(0)} &=(0,1)_6\cdot1+ (2,1)_6\cdot1+ (4,1)_6\cdot1\\&+(0,2)_6\cdot2+(2,2)_6\cdot2+(4,2)_6\cdot2\\
                                  & +(0,4)_6\cdot1+(2,4)_6\cdot1+(4,4)_6\cdot1\\&+(0,5)_6\cdot2+(2,5)_6\cdot2+(4,5)_6\cdot2
                                 ,
\end{split}\end{equation*}
\begin{equation*}\begin{split}
S(1)^{(1)} &=(2,1)_6\cdot1+ (2,2)_6\cdot2+ (2,4)_6\cdot1+(2,5)_6\cdot2\\
                                  & +2\cdot(4,1)_6\cdot1+2\cdot(4,2)_6\cdot2+2\cdot(4,4)_6\cdot1\\
                                  & +2\cdot(4,5)_6\cdot2
                                 ,
\end{split}\end{equation*}

Let $2=\gamma^b$, from the the results about the cyclotomic numbers of order 6 given in \cite{CD} which we list at the end of the paper,
we have the following cases.

From the conditions about $l$ and $q=6f+1$, we know that $f$ is odd. Then from (7), we have
\begin{equation*}\begin{split}
S(1)=&(0,1)_6+(0,2)_6\cdot2+(4,0)_6\cdot2+\\
                                  & (0,4)_6\cdot1+(2,0)_6\cdot1+(2,5)_6\cdot2,
\end{split}\end{equation*}
\begin{equation*}\begin{split}
S(1)^{(1)}=&(2,1)_6+(1,0)_6 \cdot 2+(1,2)_6 \cdot 1+(0,1)_6 \cdot 2\\
                                  & +(0,5)_6 \cdot 2+(1,2)_6+(1,1)_6 \cdot 2+(2,1)_6.
\end{split}\end{equation*}

According to the cyclotomic number of order 6 listed below, we have\\
Case I a.  $b\equiv0\mod3$:
$S(1)= -B,\ S(1)^{(1)} =(1-A)/3.$\\
Case I b.  $b\equiv1\mod3$:
$S(1) =-B,\ S(1)^{(1)} =(1-A)/3-B.$\\
Case I c.  $b\equiv2\mod3$:
$S(1) =-B, \ S(1)^{(1)} =(1-A)/3+B.$

On the basis of the cases above, combining with Proposition 1, we prove the result.
\end{proof}

\par\textbf{Example 1} Let $l=711$. Then we have $r=237$ which satisfies the conditions of Theorem 2. From $q=6f+1=A^2+3B^2$, we know $A=10$ and $B\equiv 0\bmod 3$. Then according to Theorem 2, $LC_1(\{s_n\})=LC(\{s_n\}).$

\dse{3~~Conclusion}
The $k$-error linear complexity of a sequence is an important index in cryptographic. Firstly, we give a general lower bound for the $k$-error linear complexity of subsequences of the $d$-ary Sidel'nikov sequences over the prime field  $\mathbb{F}_{d}$. Secondly, we determine the $k$-error linear complexity of subsequences of the $d$-ary Sidel'nikov sequences over the prime field  $\mathbb{F}_{d}$.

$$  \textrm Appendix$$
Cyclotomic number of order 6

Let $q$ be a prime power and $q=6f+1=A^2+3B^2$ with $A\equiv1\bmod3$ and moreover $\gcd(A,q)=1$ when $q=p^m$ and $p\equiv1\bmod6$.
Let $\gamma^b=2$, where $\gamma$ is a primitive element of $\mathbb{F}_{q}$.

Case Ia: $q\equiv7\mod12$, $b\equiv0\bmod3$

$(0,1)_6=(0,2)_6=(q+1-2A+12B)/36$,  $(0,4)_6=(0,5)_6=(q+1-2A-12B)/36$,  $(1,0)_6=(q-5+4A+6B)/36$,
$(1,1)_6=(q-5+4A-6B)/36$, $(1,2)_6=(2,1)_6=(q+1-2A)/36$.

Case Ib: $q\equiv7\mod12$, $b\equiv1\bmod3$

$(0,1)_6=(1,2)_6=(q+1+4A)/36$, $(0,2)_6=(q+1-2A+12B)/36$, $(0,4)_6=(2,1)_6=(q+1-8A-12B)/36$, $(0,5)_6=(q+1-2A+12B)/36$, $(1,0)_6=(q-5-2A+6B)/36,$

Case Ic: $q\equiv7\mod12$, $b\equiv2\bmod3$

$(0,1)_6=(0,4)_6=(q+1-2A-12B)/36$, $(0,2)_6=(2,1)_6=(q+1-8A+12B)/36$, $(0,5)_6=(1,2)_6=(q+1+4A)/36$,
$(1,0)_6=(q-5+4A+6B)/36$, $(1,1)_6=(q-5-2A-6B)/36$.


\begin{thebibliography}{99}
\bibitem{AM} H. Aly, W. Meidl, On the linear complexity and $k$-error linear com-plexity over $\mathbb{F}_{p}$
of the $d$-ary Sidel'nikov sequence, IEEE Trans. Inform.
Theory vol. 53, no. 12, pp. 4755-4761, 2007.
\bibitem{BM} N. Brandst$\ddot{a}$tter, W. Meidl, On the linear complexity of Sidel'nikov
sequences over $\mathbb{F}_{d}$, Lecture Notes in Comput. Sci., vol. 4086, Springer-Verlag, Berlin,
Heidelberg, 2006, pp. 47-60.
\bibitem{BW}  N. Brandst$\ddot{a}$tter, A. Winterhof, Subsequences of Sidel'nikov sequences,
Contemp. Math., vol. 461, 2008, pp. 33-46.
\bibitem{CD} T. W. Cusick, C. Ding,  A. Renvall, Stream ciphers and number theory, North-Holland Mathematical Library, vol. 55, North-Holland Publishing Co., Amsterdam, 1998.
\bibitem{D} C. Ding, Lower bounds on the weight complexity of cascaded binary sequences, in Adv. Cryptol..
New York: Springer-Verlag, 1991, vol. 453, Lecture Notes in Computer Science, pp. 39-43.
\bibitem{DX} C. Ding, G. Xiao, and W. Shan, The stability theory of stream ciphers, Lecture Notes in Computer
Science. Berlin, Germany: Springer, 1991, vol. 561.
\bibitem{GA} M. Z. Garaev, F. Luca, I.E. Shparlinski, A. Winterhof, On the linear
complexity over $\mathbb{F}_{p}$
of Sidel'nikov sequences, IEEE Trans. Inform. Theory, vol. 52, no. 7, pp. 3299-3304, 2006.
\bibitem{GR} A. Granville, Arithmetic properties of binomial coefficients, I. Binomial coefficients modulo prime powers,
In B. C. Burnaby (ed.), Organic mathematics 1995, CMS Conf. Proc. 20, Amer. Math. Soc. Providence, R I, (1997) pp. 253-276.
\bibitem{H} T. Helleseth, S.-H. Kim, J.-S. No, Linear complexity over $\mathbb{F}_{p}$
and trace representation of Lempel-Cohn-Eastman sequences, IEEE Trans. Inform. Theory vol. 49, no. 6, pp. 1548-1552, 2003.
\bibitem{K} Y.-S. Kim, J.-S. Chung, J.-S. No and H. Chung. On the autocorrelation distributions of Sidel'nikov
Sequences, IEEE Trans. Inform. Theory, vol. 51, no. 9, pp. 3303-3307, 2005.
\bibitem{LE} A. Lempel, M. Cohn, and W. L. Eastman, A class of balanced binary sequences with optimal autocorrelation properties,
IEEE Trans. Inform. Theory, vol. IT-23, no. 1, pp. 38-42, 1977.
\bibitem{LI} R. Lidl, H. Niederreiter, Finite fields, Second ed., Cambridge University Press, Cambridge, 1997.
\bibitem{R} R. A. Ruppel, Analysis and Design of Stream Ciphers, Berlin, Germany: Spriger-Verlag, 1986.
\bibitem{SI} V. M. Sidel'nikov, Some $k$-valued pseudo-random sequences  and nearly equidistant codes, Problemy
Peredachi Inform., vol. 5, pp. 16-22, 1969.
\bibitem{ST} M. Stamp and C. F. Martin, An algorithm for the $k$-error linear complexity
of binary sequences with period $2^n$, IEEE Trans. Inf. Theory, vol. 39, no.4, pp. 1389-1401, 1993.












\end{thebibliography}
\end{document}